\documentclass[12pt]{emulateapj}

\newcommand{\sub}[1]{\ensuremath{_{\mbox{\scriptsize#1}}}}

\shorttitle{\sc Models of AB Aurigae}
\shortauthors{\sc Jang-Condell \& Kuchner}
\slugcomment{Accepted to ApJ Letters March 19, 2010}

\begin{document}

\title{Radiative Transfer Models of a Possible Planet in the AB Aurigae Disk}
\author{Hannah Jang-Condell\altaffilmark{1,}\altaffilmark{2,}\altaffilmark{3} and Marc J.
Kuchner\altaffilmark{2}}
\altaffiltext{1}{Department of Astronomy, University of Maryland, College
Park, MD 20742, U.S.A.;
hannah@astro.umd.edu}
\altaffiltext{2}{NASA Goddard Space Flight Center, Greenbelt, MD 20771}
\altaffiltext{3}{Michelson Fellow}

\begin{abstract}

Recent coronagraphic imaging of the AB Aurigae disk has revealed a 
region of low polarized scattered light 
suggestive of perturbations from a planet at a radius of $\sim 100$ AU.
We model this darkened region using our fully non-plane-parallel 
radiative-transfer
code combined with a simple hydrostatic equilibirum approximation to
self-consistently solve for the structure of the disk surface as seen in
scattered light.   By comparing the observations to our models, we 
find that the observations are consistent with the absence of a planet, 
with an upper limit of 1 Jupiter mass.

\end{abstract}


\section{Introduction}

The Herbig Ae star, AB Aurigae (AB Aur), hosts a much-studied disk of gas and
dust thought to be representative of protoplanetary disks during 
the Jovian planet formation phase.
The circumstellar material around AB Aur has been imaged 
in scattered light by several observatories,
including STIS on HST \citep{1999Grady_etal}, 
Subaru \citep{2004Fukagawa_etal}, and 
the Lyot Project \citep{Oppe08}.  
These observations show that AB Aur is surrounded by what appears 
to be a nearly face-on disk, with structures that look like 
spiral arms.  

Using adaptive-optics coronagraphy and polarimetry, \citet{Oppe08}
imaged the AB Aur disk between radii of $43-302$ AU. 
Their images revealed an azimuthal gap in polarized light 
at a radius of about 100 AU, along with a 2-$\sigma$ bright ``spot.''
They interpreted this to indicate the presence of a possible 
massive planet of $5-37$ Jupiter masses ($M_J$).  

The position angle (PA) of the putative planet is coincident with 
the minor axis of the disk, as a well as a gap between two 
spiral arms observed in other scattered light images 
\citep{2004Fukagawa_etal},
suggesting that the darkened region results from anisotropies 
in the overall disk structure, rather than locally confined to 
a single perturber in the disk.  On the other hand, the spiral 
arms are not seen by \citet{Oppe08}.  

Another alternative is that there is no true gap in the disk 
at all, but rather that the lack of scattered polarized light 
in this region is purely a geometrical effect caused by the 
inclination of the disk \citep{2009Perrin_etal}.  
In this scenario, the disk is inclined enough that the 
far edge of the disk is back scattering rather than forward 
sacttering, creating a region of lower total polarized 
intensity in the back scattering region.  This is verified by 
a lower polarization fraction in the supposed gap, but 
no corresponding decrease in the total scattered light.  
This is likely to be the correct interpretation of the AB Aur 
observation.  The objective of this paper is to put firm 
limits on the mass of any planet embedded in AB Aur, 
complemetary to the interpretation by \citet{2009Perrin_etal}.

Planets embedded in optically thick accretion disks, like the disk
around AB Aurigae, are expected to produce perturbations in the density
and temperature structure of the disk.  \citet{paper1} and \citet{paper2}
calculated the magnitudes of these perturbations for a range of planet
masses and distances. They predicted the formation of a shadow at the
position of the planet paired with a brightening just beyond the shadow.  
\citet[][henceforth Paper I]{HJC_model} 
improved upon these calculations, by self-consistently
calculating the temperature and density structures under the assumption of
hydrostatic equilibrium and taking the full three-dimensional shape of the
disk into account rather than assuming a plane-parallel disk.

\citet{Oppe08} suggested that the observed structure resembled 
models of dust trapped in mean motion resonances with a planet 
\citep[e.g.][]{kuch03}.   However, AB Aur hosts a gas-dominated 
disk, as confirmed by detections of numerous molecular species 
\citep{1997ApJ...490..792M,2000ApJ...529..391M,2005MNRAS.359..663D,2005Pietu_etal,2005Corder_etal,2008A&A...491..821S}
In such a disk, the dynamics of the gas dominate the system 
and the dust traces the gas on orbital timescales.  
Moreover, the disk is optically thick, so the scattered light 
image traces only features in the upper surface of the disk.  
Thus, the darkened region and bright ``spot'' imaged by \citet{Oppe08} 
are unlikely to trace dust concentrations, 
as might be the case were the disk optically thin, but 
could perhaps be scattering off structure in the disk created 
in the wake of a planet.  

Here we model the observed
structure as a shadow caused by the tug of a planet in approximate
hydrostatic equilibrium with the gas disk, rather than resonant trapping
of ballistic grains.  
This interpretation is consistent with short stopping time of the dust
grains in the gas, $\sim10^{-3}$ orbits
\citep{YoudinChiang}.  We use the algorithms and code
described in Paper I
and \citet[][henceforth Paper II]{2009HJC}
to model the AB Aurigae system \citep[stellar
mass $2.4\pm0.2\,M_{\odot}$, luminosity 48 $L_\odot$, 
accretion rate $10^{-7}\,M_{\sun}\mbox{yr}^{-1}$;][]{garc06,2001Rodgers}.   
We synthesize images of the disk in scattered light and compare them with the
coronagraphic images of  the AB Aurigae disk to constrain the mass of the
perturbing planet.

\section{Model Description}

% \subsection{Model Parameters}

The model we adopt is described in detail in Papers I and II, with 
parameters adjusted for the AB Aur system.  
The stellar parameters we adopt are 
mass $M_*=2.4\,M_{\sun}$, 
radius $R_*=2.4\,R_{\sun}$, and 
effective temperature $T\sub{eff}=10^4$ K, 
which are consistent with $L*=48\,L_\odot$.
For the disk model, we assume a viscosity parameter of $\alpha=0.01$
and accretion rate $\dot{M}= 10^{-7}\,M_{\sun}\mbox{yr}^{-1}$.  
Modifying these last two parameters effectively changes only 
the overall mass of the disk.  Since we are only examining 
the perturbative effects of hypothetical planets in the disk, 
changing $\alpha$ or $\dot{M}$ do not significantly alter 
our final conclusions.  Indeed, adopting 
$\dot{M}= 10^{-8}\,M_{\sun}\mbox{yr}^{-1}$
was not found to alter our overall results.  

% \subsection{Opacities}

The amount of heating from stellar irradiation in the disk 
is sensitive to the opacities used to calculate radiative 
transfer in the disk.  
Dust opacities are perhaps the least observationally constrained 
parameter for disk modeling.  Hence, we calculate 
the opacities using a Mie scattering code developed by
\citet{1980PollackCuzzi}.  The composition of the dust is 
that used in \citet{pollack_dust}, a mixture of 
water, troilite, astronomical silicates, and organics.  
We adopt a size distribution for the dust of 
of $n(a)\propto\/a^{-3.5}$ where $a$ is the radius of the grain, 
with maximum/minimum radii of 1 mm/0.005 microns.
The large maximum grain size represents coagulation of grains 
in protoplanetary disks.  
The dust is well-mixed with the gas 
and uniform in mass opacity throughout the disk. 
The mean absorption and total extinction (absorption$+$scattering) 
averaged over a blackbody of 10,000 K representing the 
stellar spectrum are 
$\kappa_P^*=2.68$ and 
$\chi_P^*=17.3$ cm$^2$-g$^{-1}$, respectively.  
For radiation at disk temperatures, the relevant opacities are 
the flux-averaged absorption and Rosseland mean opacity, 
$\kappa_R=2.73$ and  
$\kappa_P=1.46$ cm$^2$-g$^{-1}$, respectively,
which are calculated for a blackbody at 100 K.
The temperature at the surface of the disk in the resulting models 
ranges from 64 to 143 K over the computation domain, so a 
choice of 100 K is a reasonable compromise.  

% \subsection{Scattered Light Imaging}

Scattered light dominates the surface brightness of the disk 
at 1.64 $\mu$m.  
We calculate scattered light as described in Paper II.  
The idealized images have been convolved by a Gaussian PSF 
with FWHM 0.094\arcsec, the diffraction-limited resolution of 
the AEOS telescope, to compare to the Lyot Project 
images of AB Aur \citep{Oppe08}.

% \subsection{Polarization}

To compare our model to the \citet{Oppe08} 
$P$ image in scattered polarized light, we must include the 
effects of polarization.
The size distribution of the dust grains is 
of $n(a)\propto\/a^{-3.5}$, so scattering is dominated 
by the smallest grains.  Thus, Rayleigh scattering 
is a reasonable approximation 
for the angle-dependent polarization caused by scattering
from dust grains \citep[e.g.][]{2007GrahamKalasMatthews}.
The fractional linear polarization from Rayleigh scattering is 
\citep{1974HansenTravis} 
\begin{eqnarray}
\frac{P}{I} &=& \frac{(Q^2 + U^2)^{1/2}}{I} \\
&=& 
\frac{\sin^2 \theta}{1 + \cos^2 \theta}
\end{eqnarray}
where $\theta$ is the angle of deflection. 
We assume that circular polarization is negligible, so $V=0$.  
Here, $I$, $Q$, $U$, and $V$ are the canonical Stokes parameters.  

In our model, the scatterers are dust grains in the layer of the disk
that is optically thin to stellar light.  We assume isotropic single
scattering in this optically thin layer of the disk to model the
observations.  Although our model incorporates thermalization caused
by multiple scattering of stellar photons, it does not include the
effect of multiple scattering on the polarization of the scattered
flux.  Monte Carlo calculations of multiple scattering in
circumstellar disks \citep{1996ApJ...461..847W} suggest that multiple
scattering can increase the net polarization from a smooth disk by a
factor of 1.5 to 2.  But the issue that is important to our study is
not the net polarization, it is whether the dimple caused by a planet
would be enhanced by multiple scattering effects; we will leave this
second-order effect to a future study.

\section{Comparing the Models to the Data}

The simulated images of the model disk depend on the 
assumed inclination and orientation.  This has been 
variously reported for AB Aur as 
$12-30\degr$ with PA $60-80\degr$ \citep{2006Marinas_etal}, 
$25-35\degr$ with PA $50-60\degr$ \citep{2004Fukagawa_etal}, 
$23-43\degr$ with PA $58-63\degr$ \citep{2005Pietu_etal},
and $21\degr$ with PA $59\degr$ \citep{2005Corder_etal}.  
Suffice it to say that the inclination of the disk is 
highly uncertain, although the PA is largely agreed to be 
around $60\degr$.  
We adopt an inclination of $25\degr$ as 
being a reasonable compromise between the various observations.  
For simplicity, we assume that 
the PA of the supposed planet is equal 
to the PA of the minor axis of the inclined disk.

\begin{figure*}[htbp]
\plotone{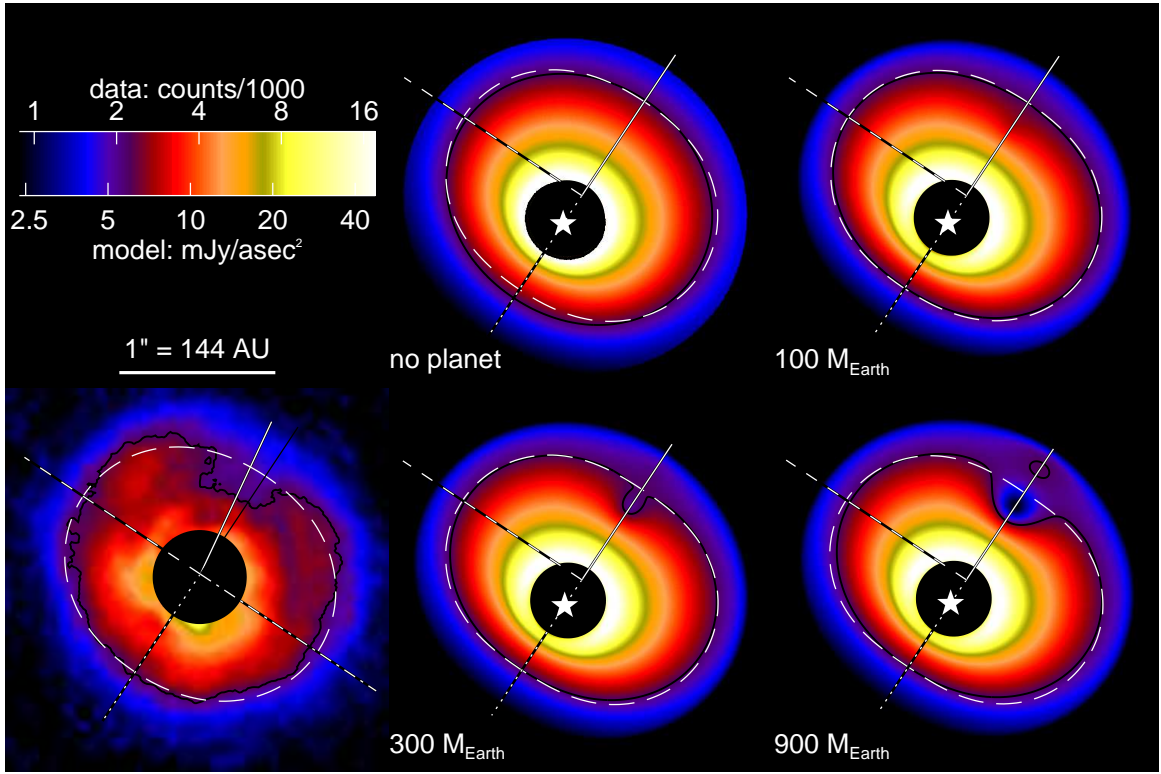}
\caption{\label{diskaxes}
Observed $P$ image and simulated images of AB Aur, scaled to the same 
spatial resolution and dynamic range.  
The observed image is in the lower left.
The blacked out inner circle represents 
the coronograph in the observed image, 
and the simulation boundary in the model images.  
The disk models contain planets of 
0 (top center), 100 (top right), 300 (bottom center), 
and 900 (bottom right) M$_{\oplus}$.  
The long-dashed while lines show the best-fitting ellipse 
to the isophotes indicated by black lines.  
The solid lines are drawn from the center of the ellipse through 
the position of the planet, which is equal to the minor 
axis in the model images.  The minor axis is indicated by a black line 
in the observed image.  
The dotted white lines show the minor axis opposite the planet.  
The dashed and dot-dashed lines indicate the major axes of the 
ellipse.  
}
\end{figure*}

Fig.~\ref{diskaxes} shows the real and simulated $P$ 
images of AB Aur in scattered polarized light at 1.64 microns, 
with the observed image from \citet{Oppe08}
in the lower left.  
We examine disk models with planets of 
0, 100, 300, and 900 $M_{\oplus}$ 
(0, 0.33, 1, and 3 Jupiter masses), as indicated.  
The color scale has been calibrated to show 
the same dynamic range in both the data and the model images.  
The blacked out inner circle represents 
the coronograph in the observed image, 
and the simulation boundary in the model images.
The northwest edge of the disk tilted away from the observer.  

There is some amount of uncertainty as to the exact location of 
the star on the observed image, 
while in the simulated models 
the stellar position is known exactly.  
Indeed, \citet{Oppe08} report an offset in the photocenter of the inner 
disk of about 88 mas or 12 AU along the minor axis.  
In order to make a fair comparison between the data and model, 
we recenter the images by the following procedure.  

We select an isophotal contour, with the brightness level set by 
the bright ``spot'' in the data image.  Although we will 
henceforth refer to this position as the ``planet,'' 
we understand it to represent an image feature 
that may or not correspond to an actual planet.  
For the model images, 
we determine the local maximum in brightness in the dimple caused by 
the 1 Jupiter mass planet, and use this brightness level across all 
model images. 
The isophotes are indicated in Fig.~\ref{diskaxes} by solid black 
lines.  We then calculate the best fit ellipse to this isophote, indicated 
by the long-dashed white lines.  The remaining calculations are now 
all calculated using the center of the ellipse as the image center.  
The calculated PA for the data is $55\degr$, while the 
planet's PA is $-25\degr$.  The angle between minor axis and 
planet PA is small, so aligning the planet with the inclination 
vector of the model disks is a reasonable assumption.  

The major and minor axes of the best fitting ellipses for the 
observed and simulated images are tabulated in Table \ref{ellipsetable}, 
assuming that $1\arcsec = 144$ AU\@.  
The offset listed for the model images is the distance 
between the star's position and the center of the ellipse.  
The best-fit ellipses for the model images are slightly more 
eccentric than that for the data, indicating that either the 
inclination of $25\degr$ is too high, or that the scattering 
properties are incompletely modeled.  For example, a small amount 
of forward scattering or slight flattening of the flared disk 
structure could both create a less eccentric best fit ellipse.  
However, we are interested in studying the perturbative effect of a
planet on the overall disk structure, not creating an exact
match to all the disk properties.  
Indeed, changing parameters such as disk inclination, accretion rate, 
and disk mass do not significantly change our overall results 
on the upper limit for the mass of a planet.  

\begin{table}
\caption{\label{ellipsetable}
Best-fit ellipse parameters to isophotal contour, in AU.
}
\begin{tabular}{lcccc}
                 & major axis & minor axis & offset & $\chi_{\nu}^2$ \\
\hline
data             &       129  &     108  &    ---   &  --- \\ 
model: \\
\quad no planet        &       131  &     104  &     31   &  0.76 \\
\quad 100 M$_{\oplus}$ &       131  &     103  &     31   &  0.44 \\ 
\quad 300 M$_{\oplus}$ &       132  &     100  &     29   &  2.18 \\ 
\quad 900 M$_{\oplus}$ &       133  &      97  &     26   &  16.8 \\ 
\end{tabular}
\end{table}

In Fig.~\ref{diskaxes}, 
the white solid, dashed, dotted, and dash-dotted lines mark the  
line through the planet, the northeast major axis, the southeast 
minor axis, and southwest major axis, respectively.  
In the simulated images, the planet is aligned with 
the minor axis, and only one major axis cut is displayed because 
of symmetry.  In the observed image, the minor axis is marked by 
a black line in the lower left of Fig.~\ref{diskaxes}, 
offset from the solid white line through the planet.  

\begin{figure*}[htbp]
\plotone{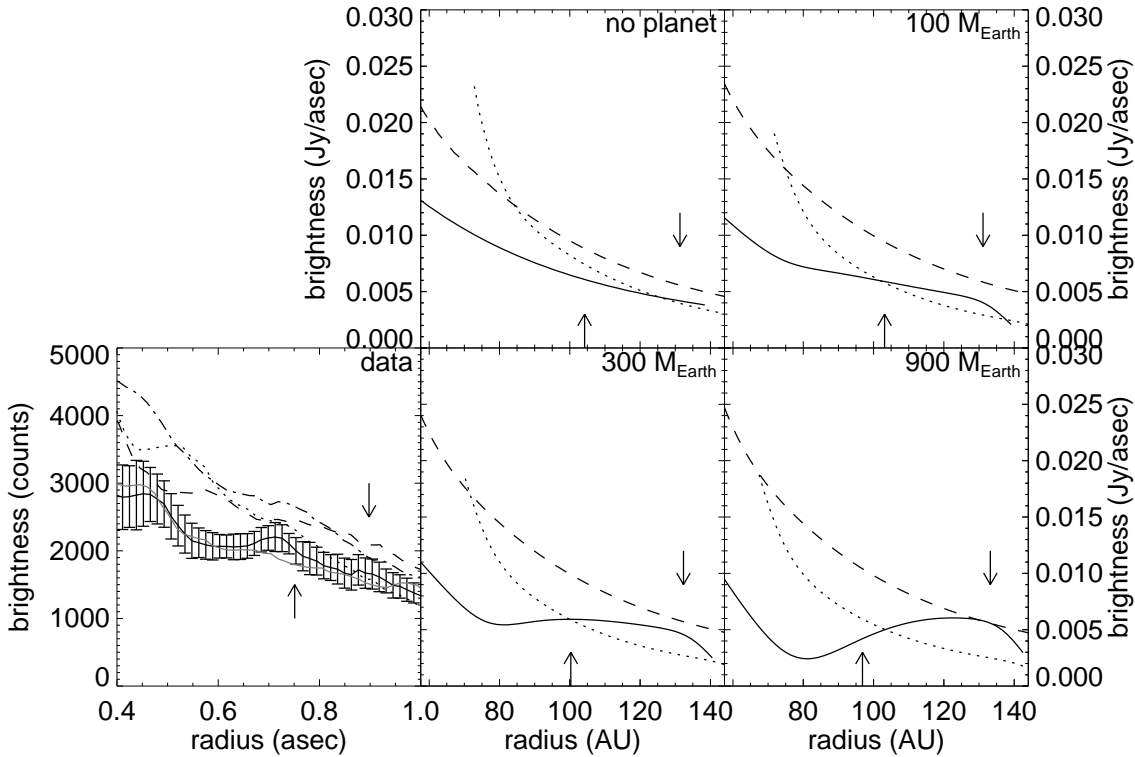}
\caption{\label{allprofiles}
Surface brightness profiles along different cuts through the 
observed and simulated disk images.  
The observed data are plotted in the lower left, 
while the remaining plots show the profiles for models with 
planets of 
0 (top center), 100 (top right), 300 (bottom center), 
and 900 (bottom right) M$_{\oplus}$.  
The lengths of the major and minor axes if the isophotal ellipse 
are indicated by 
down and up arrows, respectively.  
The solid lines are the brightness profiles 
from the center of the ellipse through 
the position of the planet.
The grey line in the lower left plot is the profile 
along the minor axis of the isophotal ellipse.  
The dotted lines are profiles along the minor axis opposite the planet.  
The dashed and dot-dashed lines are profiles along the 
major axes of the ellipse.  
}
\end{figure*}

In Fig.~\ref{allprofiles}, we plot the surface brightness profiles 
along the cuts indicated in for the observed and simulated images, 
with the solid, dotted, dashed, and dot-dashed lines indicating 
the profiles along the similarly marked cuts in in Fig.~\ref{diskaxes}, 
with the observed profiles in the lower left, and the 
remaining plots corresponding to the indicated planet masses.  
In the observed data, we plot an additional grey line, 
the profile along the northwest minor axis, indicated 
by the black line in Fig.~\ref{diskaxes}.  
This is to show that the profile does not deviate significantly 
from the profile along the planet line, except for the bright 
``spot'' pointed out in \citep{Oppe08}.  
The distances of the major and minor axes of the elliptical fit
to the isophotes are indicated by down and up arrows, respectively.   

We calculate the error bars according to the following procedure.  
At a given radial distance from the ellipse center, we sample 
the brightness along the circle of the given radius.  The inclination 
of the disk introduces an intrinsic variation in brightness that 
is symmetric about only on axis, so we find the best-fitting function 
of the form 
$a + b\sin(\theta+c) + d\sin(2\theta+e)$
where $\theta$ is the position angle and calculate the rms deviation 
from this best-fitting function to be the error.  
The errors are similar in magnitude to those calculated 
with a best-fitting function of $a + b\sin(\theta+c)$ so we are not 
over-fitting the data.  

The surface brightness profiles of the 
data (Fig.~\ref{allprofiles} lower left) 
are qualitatively simular to those  
of the planet-less model (Fig.~\ref{allprofiles} upper middle), 
with the steepest profile along the southern 
minor axis (dotted line), and the 
northern minor axis (solid line) consistently dimmer than 
the major axes (dashed and dot-dashed).  
This validates our assumption that the northwestern edge of the 
disk is tilted away from the observer.  

When a planet is added to the image, it creates a dimming 
in the surface brightness profile slightly inward of the 
length of the minor axis and a brightening just outward of it.  
The magnitude of this S-shaped perturbation grows with
increasing planet mass.  The shifts in the profiles along 
the other axes can be explained by 
the shift of the ellipse center as the planet mass increases.  

\section{Upper Mass Limit}

In order to establish how well the models match the data, 
we need to quantify the goodness of the fits.  
Since we are interested in the perturbations caused by the 
presence of a planet and only qualitatively interested in the 
rest of the disk structure, we re-scale the brightness profiles 
before directly comparing the data to 
the models.  For a given disk image, we first scale the radius 
to the length of the minor or major axis, as appropriate,
to derive a normalized distance.  In the data image, the elliptical 
radius along the planet's position differs from the minor axis 
by $1\%$, so it makes little difference to use the minor axis to 
scale this profile.  We then divide the brightness profile 
along the planet axis by that of the major axis.  This 
cancels out any small variations that might result from 
uncertainties in inclination angle.  In the case of the 
data image, we average between the profiles of both major axes before 
scaling.  

\begin{figure}[htbp]
\plotone{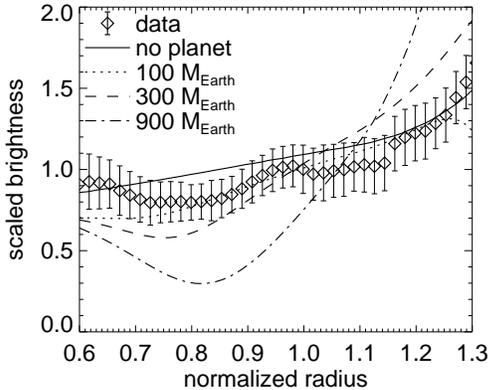}
\caption{\label{plotratios}
Scaled surface brightness profiles through the planet position.  
The data are plotted as diamonds with error bars.  
Profiles of disk models with 0, 100, 300, 900 M$_{\oplus}$
planets are 
shown with solid, dotted, dashed, and dot-dashed lines, respectively.
}
\end{figure}

The resulting normalized brightness profiles for the data and 
models are plotted in Fig.~\ref{plotratios}.  
Using these profiles and error bars measured from the data, we calculate
the reduced $\chi^2$ over the normalized radius range from 
$0.6$ to $1.3$.  
The resulting values of reduced $\chi^2_{\nu}$ are
tabulated in Table \ref{ellipsetable}.
We find that 
both a Saturn mass planet and no planet at all are good fits to the 
data, while a Jupiter mass or larger is excluded.  

\section{Conclusions}

We find that scattered polarized light images of AB Aur 
do not indicate the presence of a massive planet.
We put an upper limit of 1 $M_J$ on any planet at this location, 
well below the suggested mass of $5-37\,M_{J}$ postulated 
by \citet{Oppe08}.  Indeed, the best models are consistent with 
the absence of any planet, and support the hypothesis of 
\citet{2009Perrin_etal}, that the ``gap'' is simply an 
inclination effect rather than an indication of a planet.  

On the other hand, if a massive planet several times 
Jupiter's mass did exist in the disk, our models suggest that 
it would have been detectable.  
In other words, our work suggests that massive planets on 100 AU orbits 
can now be detected via scattered light imaging of 
nearby Herbig Ae/Be disks.  

As a Herbig Ae star, AB Aur will eventually evolve into an A-type 
main sequence star. We can compare AB Aur to two main sequence A-type 
stars with directly imaged planets:
Fomalhaut's planet is at 120 AU separation from the star 
\citep{2008Fomalhaut}, 
and HR 8799's planets are at 24, 38, and 68 AU separation \citep{2008HR8799}.  
Assuming no significant planetary migration, 
our upper limits indicate that AB Aur is not a younger analog of 
Fomalhaut.  
The coronographic spot covers the inner 40 AU 
of AB Aur's disk, so it may yet prove to be an analog of HR 8799.  
If planets do migrate significantly 
after dissipation of the gas disk, and there is some mechanism 
for expelling planets out to large distances, then AB Aur may 
evolve into a Formalhaut analog.  

If systems like Fomalhaut and HR 8799 are common and planet 
formation and migration occur early, then there may be 
many planets waiting to be detected in known Herbig Ae/Be disks.  
The requirements include 
high angular resolution ($\lesssim0.1\arcsec$), 
good star light suppression, and 
small inner working angle ($\lesssim0.3\arcsec$). 
Ongoing coronographic surveys, like the SEEDS survey on Subaru 
and future surveys on the Keck, VLT, and Gemini 
telescopes, will have the power to probe this regime.

\acknowledgments

The authors are indebted to Ben Oppenheimer for generously providing the 
observational data needed for this paper.  
We also thank an anonymous referee for helpful comments that 
greatly improved our paper.  
Simulations presented in this paper were carried out using the 
``{\tt borg}'' cluster administered by the Center for Theory and
Computation of the Department of Astronomy at the University of
Maryland.
Resources supporting this work were also provided by the NASA High-End
Computing (HEC) Program through the NASA Center for Computational
Sciences (NCCS) at Goddard Space Flight Center.
HJ-C acknowledges support from a Michelson Postdoctoral Fellowship 
under contract with the Jet Propulsion Laboratory (JPL). 
JPL is managed for NASA by the California Institute of
Technology.

\end{document}